# Fabrication of highly resistive NiO thin films for nanoelectronic applications

*Johannes Mohr*, Tyler Hennen, Daniel Bedau, Joyeeta Nag, Rainer Waser, Dirk J. Wouters*

J. Mohr, T. Hennen, R. Waser, D. J. Wouters
Institut für Werkstoffe der Elektrotechnik II, RWTH Aachen University, 52074 Aachen, Germany
E-mail: mohr@iwe.rwth-aachen.de

R. Waser
Peter Grünberg Institute, Forschungszentrum Jülich GmbH, 52428 Jülich, Germany

D. Bedau, J. Nag
Western Digital San Jose Research Center, 5601 Great Oaks Parkway, San Jose, CA 95119

Thin films of the prototypical charge transfer insulator NiO appear to be a promising material for novel nanoelectronic devices. The fabrication of the material is challenging however, and mostly a p-type semiconducting phase is reported. Here, the results of a factorial experiment are presented that allow optimization of the film properties of thin films deposited using sputtering. A cluster analysis is performed, and four main types of films are found. Among them, the desired insulating phase is identified. From this material, nanoscale devices are fabricated, which demonstrate that the results carry over to relevant length scales. Initial switching results are reported.

## 1. Introduction

Thin films of NiO see widespread use in a variety of different electronic applications, such as photovoltaic cells,[1,2] electrochromic films,[3,4] hydrogen sensors[5] and catalysts for the hydrogen evolution reaction.[6] Various forms of resistance switching, both volatile[7] and persistent,[8–12] have been observed in the material. These are believed to be promising for novel nanoelectronic devices, such as memory or selector devices in resistive crossbar arrays or for new neuromorphic hardware,[13–16] where they can emulate synapse or neuron functionalities.

Interestingly, these applications require thin films with significantly different properties. For example, a highly conductive film is desired for the hole transport layers needed in perovskite solar cells, while the film should have a high resistance state for resistive switching applications, where it can be switched to a more conductive state during operation. Depending on the exact stoichiometry, the range of resistivities reported for NiO spans approximately 10 orders of magnitude.[17] This results in different classifications of the material as either a prototypical charge-transfer insulator[18] or a p-type semiconductor.[19]

In this work, we are mainly interested in the insulating phase for applications in nanoelectronic devices, as promising switching phenomena have been observed in other correlated electron systems. [20,21] Apart from a high resistance, this requires also a precise control of other properties, such as the film density and its morphology. This is known to be challenging for NiO, of which thin films show densities less than half the bulk value,[4] and strong deviations from stoichiometry.[4] For these applications, NiO based devices must be integrated with other components such as CMOS transistors on common Si wafers, so no special substrates for epitaxial growth can be used. Additionally, processing temperatures should be compatible with standard back-end-of-line CMOS processes, about 450 °C. Thus, it must be investigated if and how the desired film can be produced under these circumstances,

In this study, the NiO thin films were grown using RF magnetron sputtering, as this is a versatile technique that allows for precise control of film properties such as the stoichiometry and morphology.

To reach a definitive conclusion on what kind of films can be produced, a large number of deposition parameters must be investigated over a wide range, without introducing a-priori beliefs about which might be important. Additionally, interactions between these parameters must be identified. This requires a carefully planned study to limit the experimental effort and assure the validity of the conclusions. For this reason, we applied statistical design of experiments.

Since it is known that parameter settings often do not transfer well from one deposition setup to another, we focus on the main trends, to determine the most important drivers of the film properties. We give special attention to the interactions between these parameters, which are often overlooked, while they can prove to be critical.

Further, a cluster analysis is performed based on the structural and electrical properties of the films to identify the main classes that can be produced. Together with the deposition data, this enables us to find a concise set of underlying parameters that determine the observable film properties.

Finally, we fabricated nanoscale devices to verify that the results obtained on full films carry over to the device sizes relevant for applications.

## 2. Results and discussion

### 2.1. Deposition and characterization

All films were deposited by RF magnetron sputtering from a NiO target, in both inert and reactive atmospheres. Oxygen and hydrogen were used as reactive gases, individually as well as in combination, to adjust the films stoichiometry. Heated depositions as well as a rapid thermal annealing were evaluated to improve crystallization. The RF power and chamber pressure, which mainly influence film morphology and stoichiometry, as well as the total gas flow during deposition were also varied. Therefore, all available settings of the system were tested.

In line with the goal of identifying key trends, a two-level factorial experiment design was selected. This tests the influence of all parameters at two levels, corresponding to a high and low setting. The main benefit of the factorial approach is that it allows to identify interactions between factors, which can lead to misleading results if only one parameter is changed at a time, as in the customary approach.[22] Considering factors at two levels is enough to elucidate if there is any significant influence, and what its direction and magnitude are. It also allows for a very efficient use of experimental effort and simplifies the evaluation of the gathered data.[23] The selected factor settings are given in **Table S1**. They are chosen to cover a large range, so effects are easily distinguished from noise, and the conclusions are relatively general.

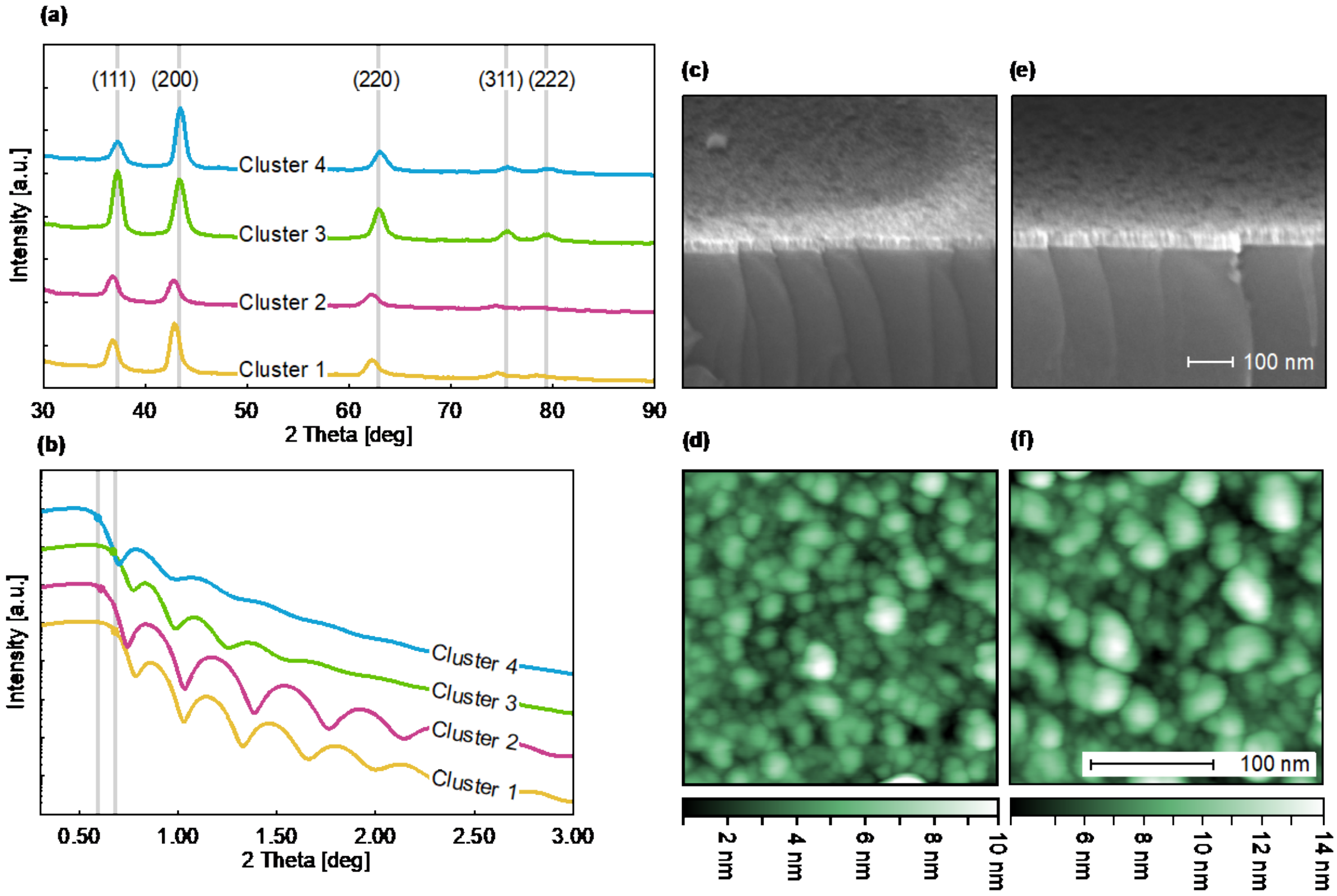

**Figure 1.** a) GIXRD results for representative films. Vertical lines indicate literature peak positions.[24] b) XRR results for representative films. Vertical lines indicate different critical angles, corresponding to low and high density films. c, d) SEM and AFM images for a low density film. e, f) SEM and AFM images for a high density film.

Typical results of the structural and morphological characterization of the films are shown in **Figure 1**. Grazing incidence x-ray diffraction (GIXRD) reveals all films to be phase pure NiO (Figure 1 a). For bulk NiO both a rhombohedral and a cubic crystal structure are reported, however the deviation from the cubic structure is so small it typically not resolved.[25] For simplicity and due to the limited signal from thin films, the cubic structure is assumed in the analysis. Below, four main types of films are identified, a representative example is shown for each one (Colors are used consistently throughout this work). They have either the lattice constant of the bulk material of 4.177,[24] or a larger value of the cubic lattice constant, due to non-stoichiometry. A significant preferential orientation is indicated by a deviation of the ratio of the (111) peak to the (200) peak from the reported value of 61%[24] of a powder sample. The combined intensity of all peaks serves as a proxy for the degree of crystallization of the films. X-ray reflectometry (XRR) measurements (Figure 1 b) confirm a film thickness of 30 nm and the growth of a homogeneous film. Approximately, two different critical angles can be identified, highlighting the significant difference in density between films. The quicker decay in the oscillations for the top two curves is due to higher interface roughness. Scanning electron microscopy (SEM) and atomic force microscopy (AFM) reveal a dense, columnar microstructure, with a fine grain structure for all films (Figure 1 c – f). The higher density film appears to be composed of fewer, but larger grains.

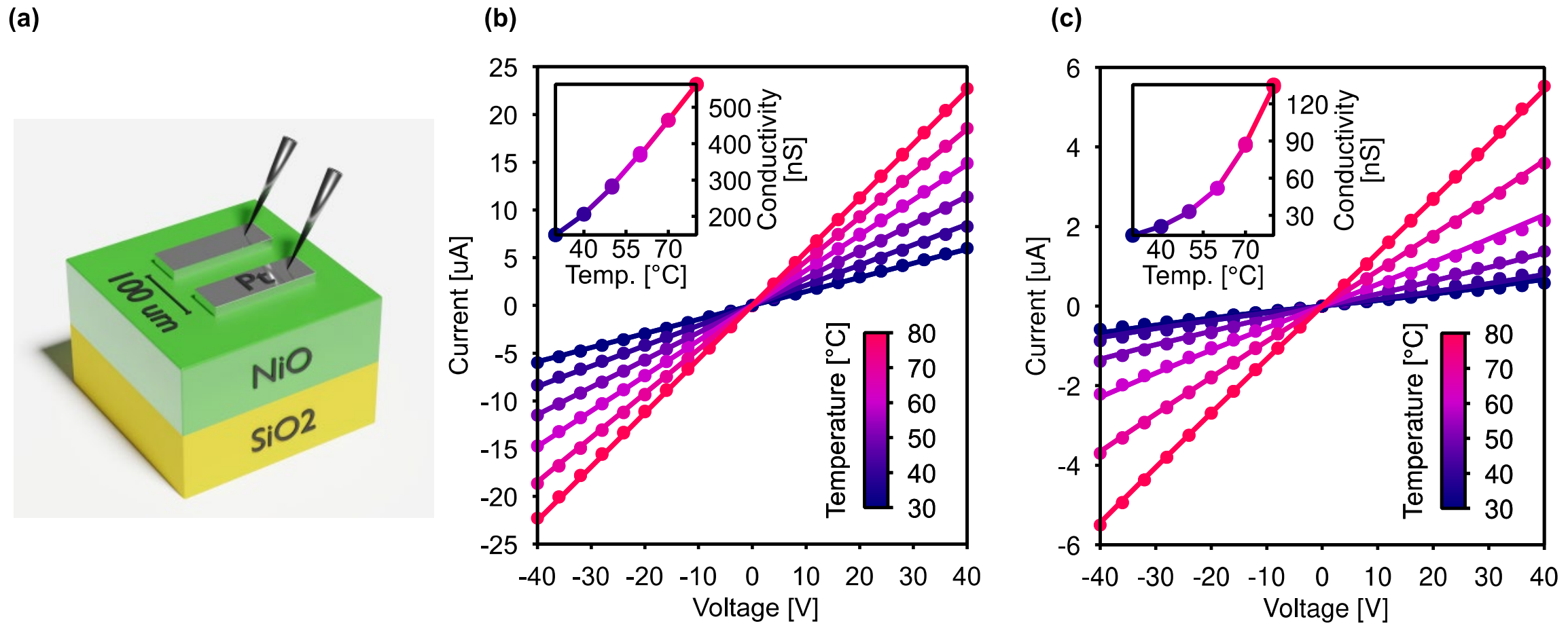

**Figure 2.** a) Geometry of the electrical measurements. b, c) IV characterization results for a sample with a conductivity linear in temperature (c) and one showing a more non-linear dependence. The conduction behavior differs between the clusters identified below, the linear one belongs to cluster 1, the non-linear one to cluster 2. Note that the dependence of the conductivity on temperature is significantly stronger in c). This result is typical, as evident from the cluster analysis.

After the X-ray-analysis, 30 nm thick Pt electrodes were deposited by DC-sputtering and structured via standard lift-off photolithography, to facilitate the characterization of the electrical properties.

The resistance of the samples was measured between two electrodes with a length of 300 µm and a separation of 100 µm (**Figure 2 a**) using a Keithley 6430 source meter. As many of the films exhibit a high resistivity, and thus a high RC constant, the resulting charging time must be taken into account and the measurement voltage (typically 5V) was applied for a soak time of 1 min before the current was measured. The procedure was repeated with inverted voltage polarity to cancel the effects of any offset currents. Due to the large range of resistances occurring, the logarithm of the actual value is reported below.

Additionally, differences in the conduction mechanisms between the films were analyzed by preforming voltage sweeps between -40 V and 40 V at various temperatures ranging from 30 °C to 80 °C. Figure 2 b) and c) shows two exemplary measurements exhibiting a different temperature dependence. To enable a quantitative comparison, the resulting IV characteristics must be summed up in a descriptive set of numbers. This is achieved by fitting a phenomenological model to the data:

$$I = aV + bV|V| + cVT + dVT^2 \qquad (1)$$

$$G = a + b|V| + cT + dT^2 \qquad (2)$$

This allows for a non-linear dependence of the conductivity on temperature, and a linear one on voltage. It was verified that the current depends only on the magnitude of voltage and not its polarity, as expected due to the symmetrical electrode configuration. Because the lateral measurement geometry limits the achieved field strengths, higher order terms in voltage can be neglected. Note that due to the conductance depending linearly on voltage, the measured current contains a squared voltage term.

The benefit of using a phenomenological description instead of a physically based conduction mechanism formula arises in the classification of the films. Because they are all described by the same model, the similarity or difference between their behaviors is easily measured. Since the effects of the determined coefficients are difficult to visualize, the total dependence on temperature (TTD), the deviation from a linear temperature dependence (NLT) and the total dependence on voltage (VD) are reported below. These are calculated from the corresponding normalized derivative of the conductivity G, evaluated for an average measurement voltage and temperature:

$$TTD = \frac{\frac{d}{dT}(a+bV+cT+dT^2)}{a+bV+cT+dT^2}_{T=55°C,\,V=20V} \qquad (3)$$

$$NLT = \frac{\frac{d^2}{dT^2}(a+bV+cT+dT^2)}{\frac{d}{dT}(a+bV+cT+dT^2)}_{T=55°C,\,V=20V} \qquad (4)$$

$$VD = \frac{\frac{d}{dV}(a+bV+cT+dT^2)}{a+bV+cT+dT^2}_{T=55°C,\,V=20V} \qquad (5)$$

They can be interpreted as the percentage change in the conductivity or its slope for a change in temperature or voltage, for average conditions.

## 2.2. Factorial analysis

For the analysis of the factorial design linear models were used. The low and high settings of a factor are represented by -1 and 1 respectively. The models have the general form

$$y = \text{intercept} + a_1 x_1 + a_2 x_2 + \cdots + a_{12} x_1 x_2 + \cdots + \text{error} \quad (6)$$

with $y$ representing the measured property and $x_i$ the deposition factors. These models were fit to the data by least squares regression using the R language.[26]

The complete models include 25 terms. For a practical understanding of the processes smaller models are preferable. Therefore, only the necessary large coefficients were included so that 90% of the variation is explained. The significance of these was estimated, and any terms found to be insignificant were removed from the model.

To gain an overview of the relation between deposition factors and material properties, the results of the modeling can be summed up in a graphical representation (**Figure 3**). The models for the properties describing the conduction behavior are not reported here, as they were found to be large and provide limited insight. These are treated in a different way below.

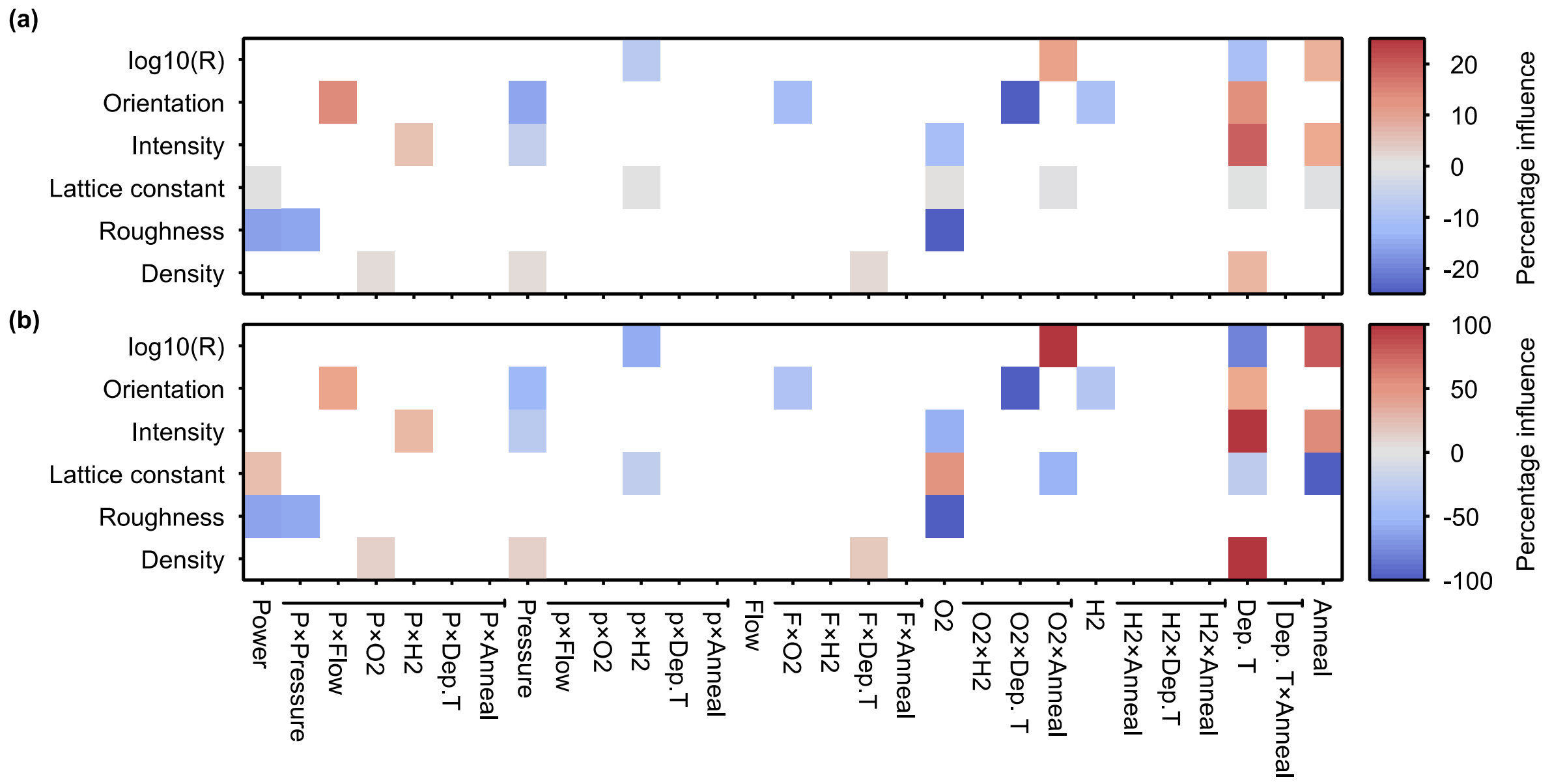

**Figure 3.** The influence of the deposition factors on various physical properties. The x-axis corresponds to all terms from **Equation 6**. The single factors are the $x_i$ terms, while the products denote the interactions $x_i \times x_j$. Here, the abbreviated first factor is the single factor to the left (e.g. P = Power, p = Pressure, etc.). Each row represents the model for the property on the y-axis. The colors give the magnitude of the coefficients $a_i$. Coefficients not included in the model are white. a) Shows the coefficients normalized by the intercept, which is the average value of that property for all samples. E.g. if the deposition temperature is set to the high value, the film density is expected to be ca. 15% above average, for the low setting 15% below. b) The same data, but normalized to the largest value. This more clearly shows the individual influences for properties that change little in absolute terms, like the lattice constant.

This allows for a quick identification of the most important drivers of the properties, and especially which properties can be tuned separately. For example, the film density is controlled by the deposition temperature, and the roughness by the power and oxygen concentration (as well as the interaction between power and pressure). This is important because both properties can be optimized for a target independently, and thus the best possible value reached for each, while for pairs of properties that have many drivers in common, such as the resistance and degree of crystallization (measured by the diffracted intensity), trade-offs might be necessary.
Numerical values for the fitted models are given in **Table S2**.

### 2.3. Cluster analysis

To discover further structure in the data, a cluster analysis was performed using the k-means clustering algorithm. This tries to identify groups of films with similar properties, which may be regarded the main types of films. For this, all measured properties were considered, those used for the factorial analysis as well as the conductivity properties. First, the right number of groups must be identified. By inspecting the amount of explained variation over the number of clusters (**Figure S3**), it can be concluded that a satisfactory partitioning is achieved by assuming four main groups in the data. To verify this, the measured data can be shown separately for each cluster (**Figure 4**).
It is seen that some properties show the clearest distinctions, those being the density, lattice constant, roughness and film resistance. However, not all properties are independent. For example, the films with a larger lattice constant are also more conductive. The likely cause is a super stoichiometric oxygen content. This is known to lead to an elongated lattice constant compared to the literature value of the stoichiometric oxide of 4.177.[24] The oxygen excess is also known to produce defects (likely Ni vacancies) which cause a p-type conductivity.[19,27] Additionally, it is seen that on average, the conductive films also grow smoother. Thus, these three properties can be seen as effects of one underlying driver, the stoichiometry.

Applying the same reasoning, it becomes clear that the films that are dense have a conduction mechanism independent of voltage, while the low-density films show a pronounced dependence on voltage. Furthermore, the dense films have strong diffraction peaks, while this is weak for the low-density clusters. The difference in diffracted intensity can be attributed to the larger size of the grains observed by AFM (Figure 1 d, f) for dense films. This corresponds to a reduced number of grain boundaries, which can be assumed to act as barriers, leading to the non-linear conduction. It is likely that the density at grain boundaries is less than in the center of the grains, which is consistent with the experimental observation that films with fewer grain boundaries are more dense. From these considerations, the films' concentration of grain boundaries can be regarded as a second principal dimension.

In Figure 4 c) the four clusters are assigned to these two orthogonal dimensions, along with the remaining independent properties. These indicate that a less pronounced third dimension exists. It can be identified by comparing the preferred orientation of the films to the temperature dependence of the conduction mechanism. The films with a peak ratio of approximately the literature value of 61% [24] (cluster 1) show nearly no temperature dependence of the conductivity, while those with a very pronounced orientation do (cluster 2+3). For the moderately ordered films of cluster 4, a temperature dependence between the oriented and the random film is found.

If this has indeed been correctly identified, it might be speculated that three orthogonal dimensions should lead to eight possible main classes of films, of which only half have been observed. The remaining ones might be accessible through alternative fabrication routes.

Considering the ways to produce a film in a specific cluster, Figure 4 b) shows that only some of the parameters are important. For the deposition power, pressure and total gas flow, the data are distributed about equally between the low and high factor settings, implying that these are not significant. On the other hand, the reactive gas contents and especially the thermal process steps, appear decisive.

The film density is mainly controlled by the deposition temperature. This becomes clear by comparing the corresponding rows in Figure 4 a) and b). For two clusters, a high temperature was used predominantly, and for both a high density was measured. For the two low density clusters, depositions were done at room temperature exclusively. This is in agreement with the results from the modeling done above.

With regard to the film resistance, the advantage of combining both analysis methods becomes clear. For the relatively conductive films, $O_2$ has to be added, and no post anneal should be performed. This is very clear from the distribution of settings favoring one level.

On the other hand, for the insulating ones, while no $O_2$ and a post anneal are somewhat preferred, the difference is far less obvious. This can be understood from the analysis of the factorial design. Because of the interaction between $O_2$ and anneal, the $O_2$ flow has an influence only when no anneal is done. If on the other hand no oxygen is added, the anneal has little effect.

This is another indication that the conduction is driven by the stoichiometry. Without added oxygen, this is likely close to the ideal value (due to the deposition from a stoichiometric target), and the film is insulating. The anneal probably leads to a thermal reduction of the material, as has been reported in [28]. Apparently, this is only the case if excess oxygen is present, a stoichiometric film is not reduced further (at this temperature).

This very useful for the fabrication of the desired insulating film because it adds a level of resilience against random process fluctuations. In addition, as seen in Figure 3, growing a smooth film requires deposition with an oxygen flow. The anneal enables the fabrication of films that are both smooth and high resistive, by removing the excess oxygen later.

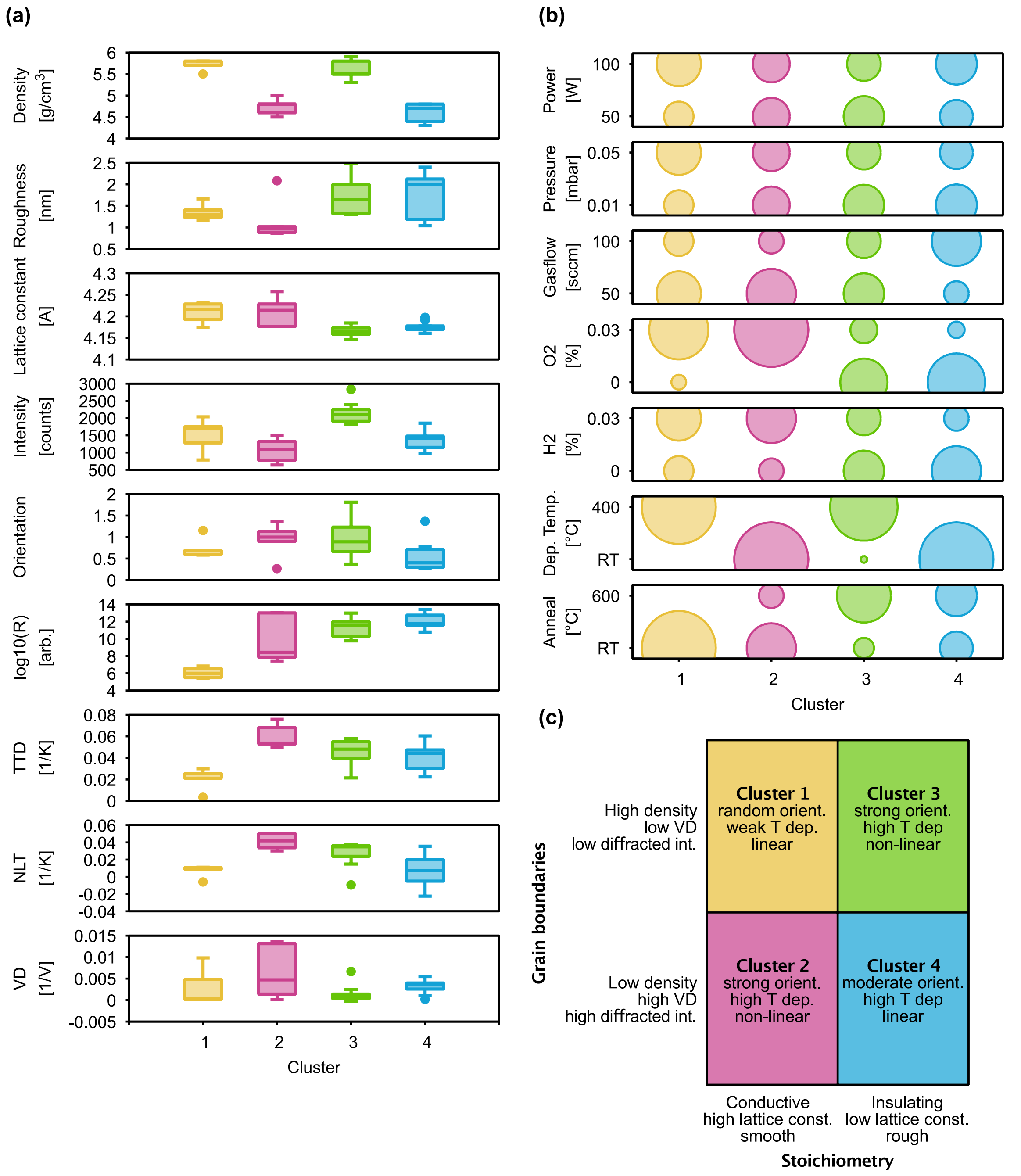

**Figure 4.** Results of the cluster analysis. a) Distribution of the properties of the film for the four clusters. b) Settings used to deposit films in a particular cluster. The size of the circles corresponds to the percentage of depositions that used that factor level. This allows to identify settings that are common to most films in the cluster (one circle is much bigger, e.g. the deposition temperature for cluster 1), and therefore important to produce a film of this type. Settings where high and low value are equally likely (e.g. power for cluster 1) are not relevant for the assignment to this cluster. c) Assignment of the clusters to two main, orthogonal dimensions.

## 2.4. Nanodevices

To verify that the conclusions obtained on full films are meaningful also at the nanoscale, for a subset of the deposition conditions, nanodevices were fabricated. The conditions were selected so that for each of the clusters identified below, two samples were made. The films were deposited on TiN via structures, with via sizes ranging from 500 nm × 500 nm down to 120 nm × 120 nm (**Figure 5 a, b**). Before processing the surface was cleaned by Ar sputtering to remove any surface oxide. A 30 nm NiO film was then deposited, and the anneal was performed where required. Finally, a 30 nm Pt top electrode was deposited by RF sputtering.

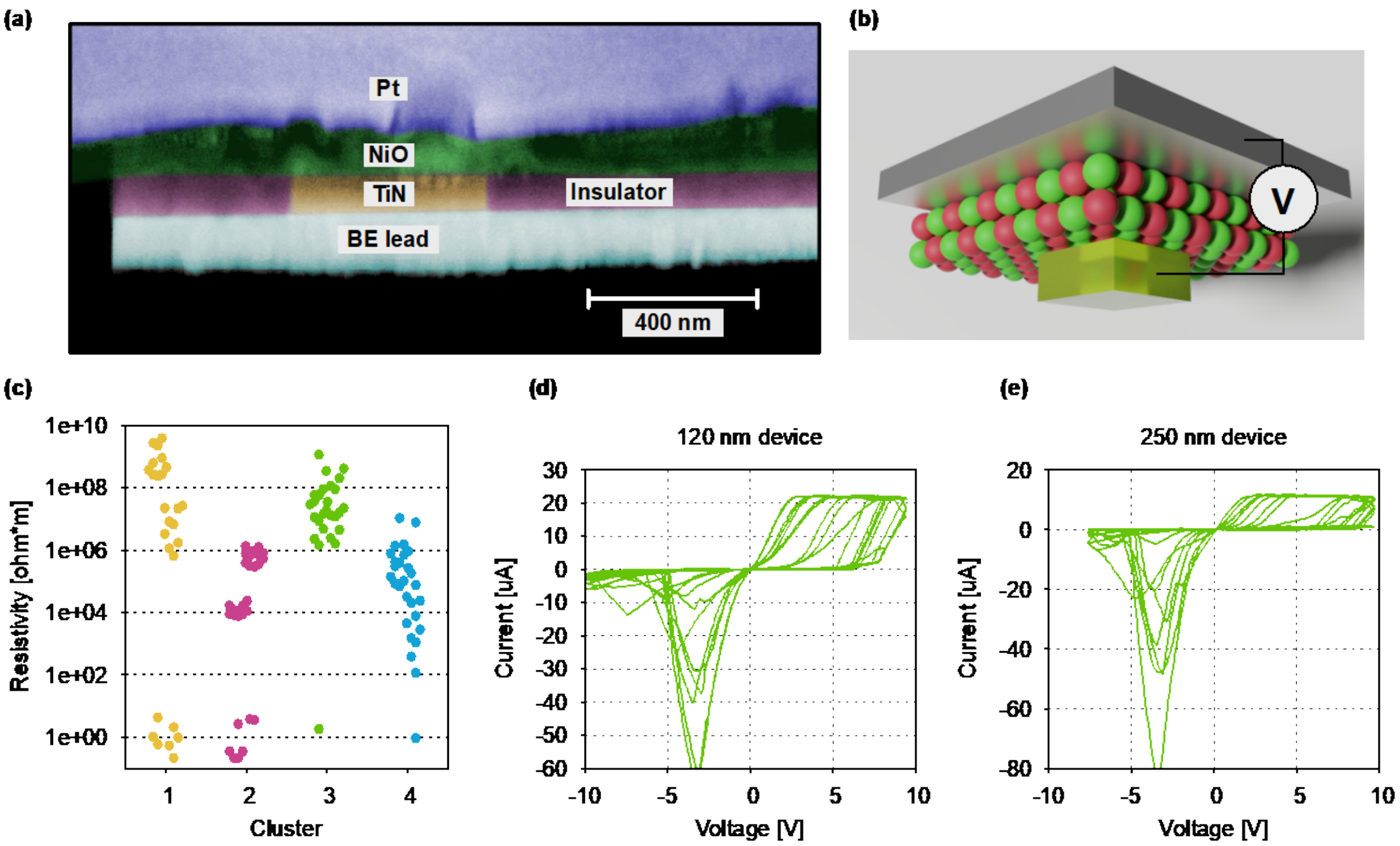

**Figure 5.** a) SEM cross section of a 500 nm wide device. b) Sketch of the measurement geometry. c) Distribution of measured device resistivities for the four clusters. d, e) Exemplary switching characteristics for two devices fabricated from cluster 3 films.

To characterize the electrical properties of the nanodevices, first the pristine film resistivity was measured, at ambient conditions, for 14 devices of various sizes on each sample. The results (Figure 5 c) indicate that for most devices a high resistivity is achieved. In the first two clusters, however, a significant number of low resistance devices are also found, and in the last two only a single outlier each. This is not surprising as the bulk measurements show these film types are more conductive. That also high resistive devices are seen is likely due to an interface layer of stoichiometric NiO on the interface of the TiN via. At the high deposition temperatures, it is possible for the TiN to be oxidized (**Figure S4** demonstrates the formation of $TiO_2$ for full films). Most likely, this is not enough to significantly influence its conductivity, but the oxygen missing from the NiO makes the film stoichiometric, and therefore insulating at the interface. Because the resistivity of the insulating material is many orders of magnitude higher, even a thin interface layer can dominate the device resistance. However, the interface layer will often be broken by defects, and the low resistivity of the remaining film is measured.

Figure 5 also suggests that cluster 2 is less resistive than cluster 1, and cluster 4 less resistive than cluster 3. At first glance this seems to contradict the results from the cluster analysis, where the opposite trend in resistances was observed. However, it was also found that for cluster 1+3, the conductivity is largely independent of the voltage amplitude, while for the other two there is a pronounced influence. Since in the nanodevices the resistance is measured across the 30 nm film, while in the full film samples between electrodes 100 µm apart, the achieved field strength is much higher. Therefore, the samples that are sensitive to the field amplitude appear to be much more conductive.

Finally, switching attempts were made on some samples using a setup specifically developed for the characterization of nanoscale RRAM cells.[29] This enables the use of a precisely controlled current compliance during the experiments. The samples for the switching experiments were selected from the film type with the most bulk like properties (cluster 3), these also exhibited the most homogeneous electrical properties in the nanodevices. A typical bipolar resistive switching behavior is found. The operating currents are remarkably low, a current compliance value of only 10 µA is demonstrated in Figure 5 e). No forming procedure was necessary to achieve a switching effect.

A detailed investigation of the switching effect is beyond the scope of this work, but since one of the films was deposited with additional oxygen flow and the other was not, it can be concluded that the finding from full films, that the post anneal removes any excess oxygen, also holds on the nanoscale.

**3. Conclusion**

Based on our analysis, the most desirable films for device applications are found among those in cluster 3. These films possess the desired high resistivity, while also being reasonably dense. Since these films also show the best degree of crystallization and a bulk-like lattice constant, they can be expected to behave most closely like ideal bulk NiO.

To prepare this film type, it is found that thermal steps are necessary, a high temperature deposition is needed to grow a dense film, and a post annealing step is required to adjust the stoichiometry. Oxygen flow during the deposition is needed to produce the smooth film required for applications.

The other deposition parameters are much less important with regard to the main properties, and can therefore be chosen based on other considerations. The power should be set high to reduce the deposition time, and a low process pressure should be used to reduce contaminations in the film. Since it appears that using $H_2$ in the process has little effect when targeting high resistive films, it should be avoided for simplicity.

The measurements on the fabricated nanodevices demonstrate that the results for full films carry over quite well. Again, two main dimensions can be identified, driving the average conductivity, which is dominated by the presence or absence of individual low resistance devices, and the degree of field activation in the conductance mechanism.

The identified resistance switching effect appears promising due to the very low current operation, which is desirable in practical devices. The switching voltages exceed the voltage range available in modern CMOS circuits, however they can likely be reduced by scaling down the film thickness. This is enabled by the high resistivity and homogeneity between devices.

## 4. Experimental Section/Methods

*Sample fabrication*: All samples were fabricated on 1” × 1” coupons diced from 150 mm Si wafers with a ~400 nm thermal $SiO_2$ layer on top. The substrates were cleaned in an ultrasonic bath in acetone and isopropanol for 10 min each, and thoroughly rinsed in deionized water afterwards. Films were deposited in a custom-built RF magnetron sputtering system from a 1” NiO target (commercially acquired from Evochem). The sputter sources are located at a 45° angle to the rotating substrate. Up to three gases were introduced into the chamber via mass flow controllers in different amounts: Ar as a background gas for the sputtering process, $O_2$ to provide an oxidizing atmosphere and $H_2$ for a reducing environment. In addition to experiments with one reactive component, an inert atmosphere Ar was considered as well as Ar + $O_2$ + $H_2$ mixtures. The total pressure in the chamber was controlled by an outflow valve. A subset of the depositions was done on substrates heated from the backside via halogen bulbs. In addition, on some samples a subsequent rapid thermal annealing step was carried out for 1 min at 600 °C in ambient pressure nitrogen, to determine if this can be a substitute for a high temperature deposition.

The deposited film thickness was controlled by the deposition time. Since it can be expected that film properties depend on thickness, all depositions were tuned to a target of 30 nm. This required an additional preliminary run to determine the deposition rate.

*X-ray analysis*: All measurements were done in a Panalytical Xpert Pro MPD diffractometer using Cu K $\alpha 1$ radiation. The film thickness, density, and interface roughnesses were determined by X-ray reflectometry (XRR) measurements. To extract these parameters, a $NiO/SiO_2$ bilayer model was fit to the data using the Panalytical X'Pert Reflectivity software. For each sample, the roughnesses for the $SiO_2/NiO$ and NiO/air interfaces were extracted, and the maximum was reported. The rougher interface was almost exclusively the top, NiO/air one. Grazing incidence diffraction (GIXRD) was measured at a 0.4° incident angle, this was optimized to yield as strong a signal from the thin-film as possible and minimize substrate influence. The data was analyzed by comparison to literature values from [24].

*Morphological characterization*: The film morphology was verified by inspecting the cross section of fractured samples using scanning electron microscopy in a Zeiss DSM 982. Atomic force microscopy images were obtained using a Park NX10 system with OMCL-AC160TS cantilevers in non-contact mode.

*Experiment design*: A $2^{7-2}$ fractional factorial design was used.[30] This requires 32 runs. The design is a resolution 4 design, therefore all main effects are resolved. In principle, some two factor interactions could be confounded, but since this is the case for only few interactions, these can be assigned to factors where an interaction is physically not reasonably expected. Therefore practically, two factor interactions are also resolved. Three factor interactions were assumed to be negligible, as is usually the case.[31]

*Cluster analysis*: Before the analysis, the data were standardized (per property) to achieve a distribution with mean 0 and standard deviation 1. Then, the k-means clustering algorithm as described in [32] (and implemented in R[26]) was applied. Since the results are affected by the selection of initial cluster centers, 200 separate runs were done with random initial selections, then the best was selected. The appropriate number of clusters was identified via the "elbow method", and found to be four.

**Acknowledgements**

The authors wish to thank S. Taranenko and Prof. M Wuttig from RWTH I. Institute of Physics for providing the FIB/SEM device image.

The authors acknowledge funding by the DFG (German Science Foundation) within the collaborative research center SFB 917.

# Supporting Information

**Table S1.** Deposition factors

| | Power | Pressure | Gasflow | $O_2$ | $H_2$ | Dep. Temp. | Anneal |
|---|---|---|---|---|---|---|---|
| High | 100 W | 0.05 mbar | 100 sccm | 3% | 3% | 400 °C | 600°C |
| Low | 50 W | 0.01 mbar | 50 sccm | 0% | 0% | RT | None |

**Table S2.** Fit models

| Property | Equation |
|---|---|
| Density [g/cm$^3$] | $5.19 + 0.49*DepT + 0.11*H_2*Anneal + 0.075*Pressure + 0.075*Power*O_2$ |
| Roughness [nm] | $1.69 - 0.44*O_2 - 0.27*Power - 0.25*Power*Pressure$ |
| Lattice constant [A] | $4.18 - 0.019*Anneal + 0.011*O_2 - 0.01*Anneal*O_2 + 0.0057*Power - 0.005*DepT - 0.0045*Pressure*H_2$ |
| Intensity [counts] | $1604 + 317*DepT + 195*Anneal - 174*O_2 + 112*Power*H_2 - 92*Pressure$ |
| Orientation [arb.] | $0.8 - 0.24*DepT*O_2 + 0.12*Power*TotalFlow - 0.12*Pressure + 0.12*DepT - 0.09*O_2*Gasflow - 0.08*H_2$ |
| Log10(R) [arb.] | $10.1 + 1.32*Anneal*O_2 + 1.07*Anneal - 1.05*DepT - 0.75*Pressure*H_2$ |

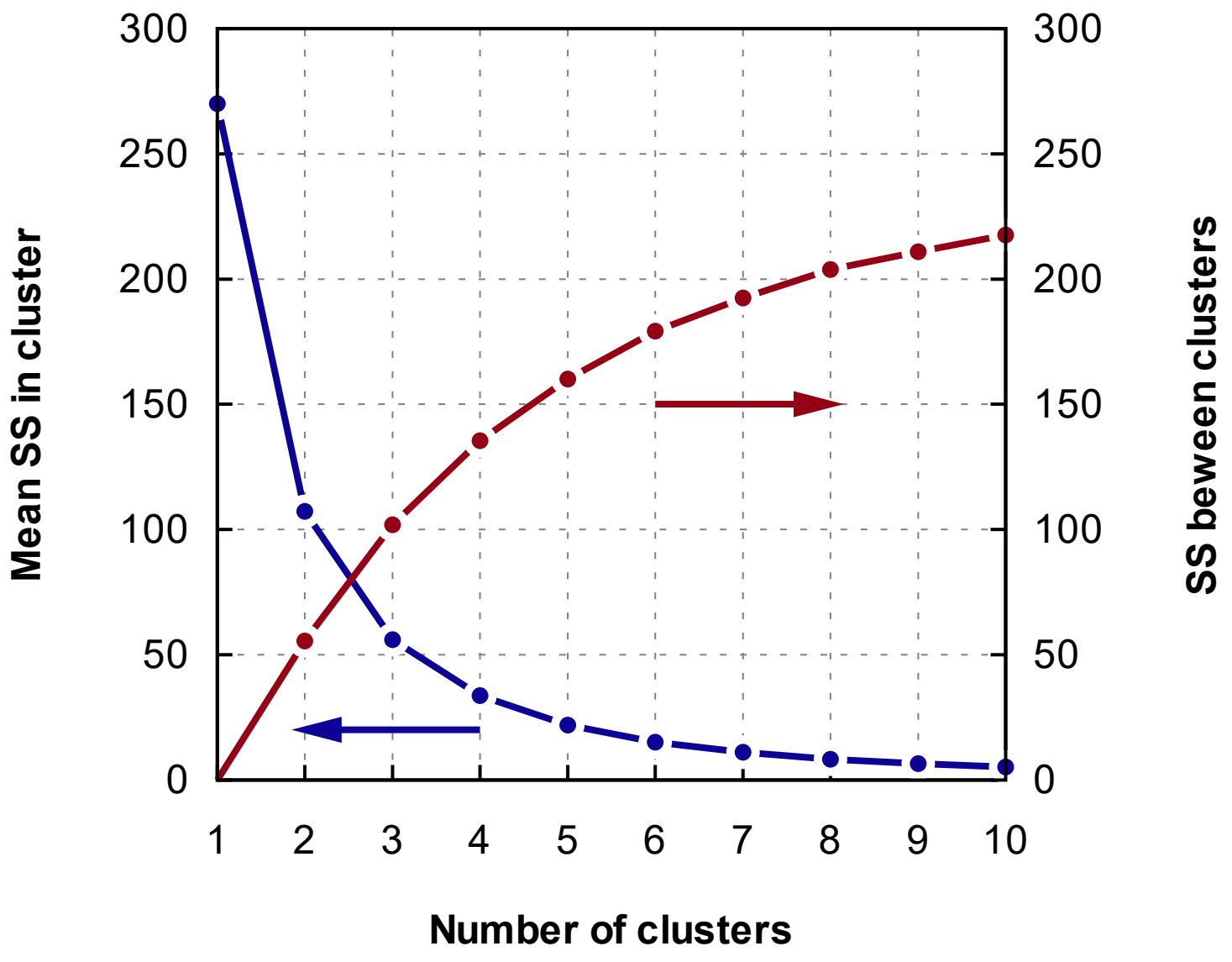

**Figure S3.** The average variation within the clusters (blue), as measured by the sum-of-squares (SS) of the deviation from the cluster center, for various numbers of clusters, as well as the variation between cluster centers (red).

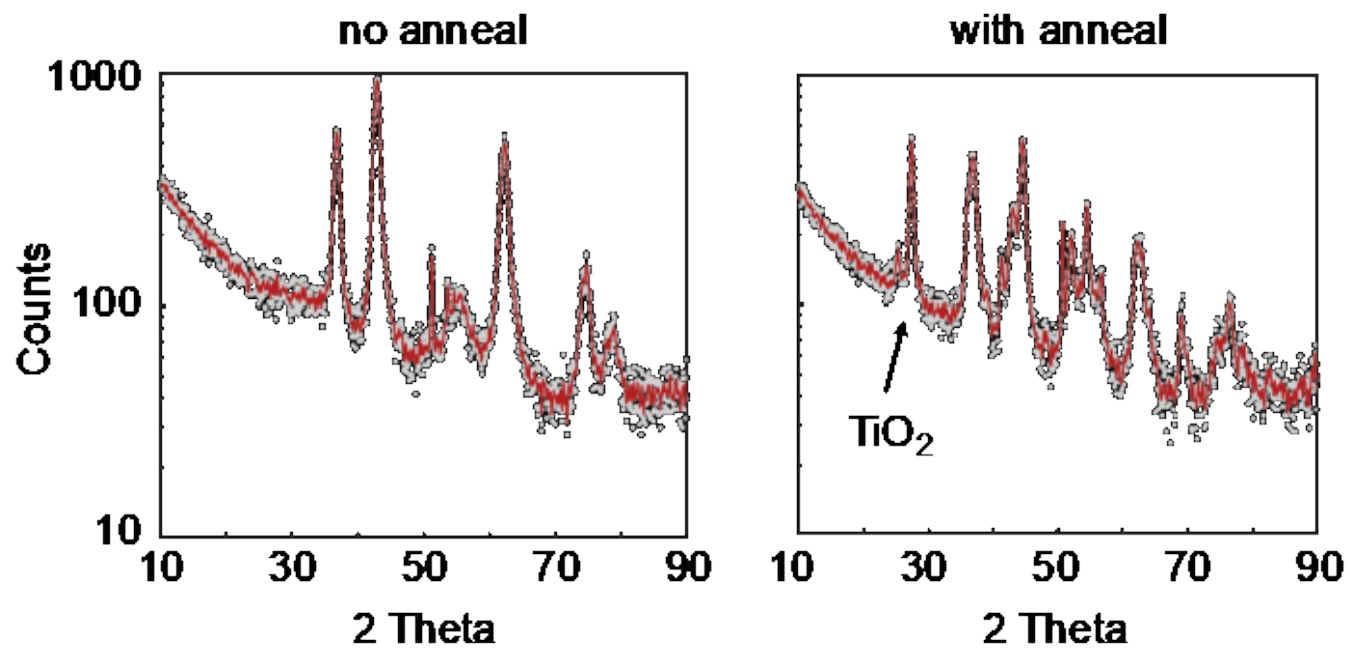

**Figure S4.** GIXRD for NiO on TiN. The left measurement shows the peaks for NiO and TiN, which are very close and hard to distinguish. This sample was not annealed, while the one on the right was processed with the RTA step discussed in the main text. The more complex peak structure demonstrates the formation of additional phases, of which $TiO_2$ can be identified. Note that the TiN used here is not identical to the material used for the vias, and is believed to be of a lower quality.